\documentclass[twocolumn,showpacs,preprintnumbers,amsmath,amssymb]{revtex4}
\usepackage{graphicx}
\usepackage{dcolumn}
\usepackage{bm}

\newcommand{\be}{\begin{equation}}
\newcommand{\ee}{\end{equation}}
\newcommand{\nc}{\newcommand}
\nc{\bi}{\bibitem}
\newcommand{\ba}{\begin{array}}
\newcommand{\ea}{\end{array}}

\def\tr{\mathrm{tr}}

\newcommand{\beq}{\begin{equation}}
\newcommand{\eeq}{\end{equation}}

\newcommand{\myref}[1]{~{(\ref{#1})}}
\newcommand{\mycite}[1]{~{\cite{#1}}}
\newcommand{\myfigref}[1]{~{Fig.~(\ref{#1})}}
\newcommand{\req}[1]{Eq.\,(\ref{#1})}

\makeatletter
\newcommand\figcaption{\def\@captype{figure}\caption}
\makeatother

\begin{document}
\preprint{ITEP-TH-19/09}
\title{The Confinement Property in SU(3) Gauge Theory}

\author{A.~V.~Zayakin$^{1,2}$}
 \email{Andrey.Zayakin@physik.lmu.de}
\author{J.~Rafelski$^{1,3}$}
  \email{rafelski@physics.arizona.edu}

\affiliation{$^1$  
Department f\"ur Physik der Ludwig-Maximillians-Universit\"at M\"unchen und\\
 Maier-Leibniz-Laboratory, Am Coulombwall 1, 85748 Garching, Germany
}
\vspace*{2mm}
\affiliation{$^2$  
ITEP, B.Cheremushkinskaya, 25, 117218, Moscow, Russia
}
\vspace*{2mm}
\affiliation{$^3$ 
 Department of Physics, University of Arizona, Tucson, Arizona, 85721 USA}

\date{May 10, 2009}

\begin{abstract}
We study confinement property of pure SU(3) gauge theory, combining in this effort the non-perturbative gluon and ghost propagators obtained as solutions of Dyson--Schwinger
equations with solutions of an integral ladder diagram summation type equation for the
Wilson loop.  We obtain  the string potential and effective UV coupling.
\end{abstract}

\pacs{12.38.Aw,11.15.Pg,11.15.Tk}

\maketitle

\section{Overview}
The problem of explaining quark confinement  has been of foremost importance since
the formulation of quantum chromodynamics (QCD). The principal manifestation of
confinement is the linear growth of QCD potential between color charges. This is known
to be the property of  the Wilson loop~\cite{Wilson:1974sk}. However, it has been
impossible so far to use this in an analytic {\it ab initio} calculation in $3+1$
dimensional QCD. We want to deal with this challenging problem by combining:\\
a)  the Erickson--Semenoff--Szabo--Zarembo (ESSZ)~\cite{Erickson:1999qv,Erickson:2000af}
 formulation for   Bethe--Salpeter type equation   for Wilson loops, with\\
b)   Dyson--Schwinger equations  (DSE) for the gluon and ghost propagator in Landau
gauge~\cite{Alkofer:2008bs,Huber:2007kc}.\\
We solve DSE for gluons and ghosts
in the pure glue two-point sector. Then we insert the resulting
QCD coupling $\alpha=g^2/4\pi$ and the gluon propagator into
ESSZ equation for a rectangular (non-supersymmetric)
 Wilson loop, and solve this integral  equation,
which yields the Wilson  potential.

The ESSZ type ladder (or rainbow) diagram summation has long been a major tool
for extracting non-perturbative information about dynamics of a gauge theory. However
the strength of this method is more evident in $\mathcal{N}=4$ supersymmetric Yang--Mils due to higher order
vertex correction cancelation. In principle the  use
of ESSZ ladder summation in our context of non-supersymmetric QCD is doubtful, and we will
make several efforts to establish this approach: we will study the vertex correction terms
 by comparing the leading order (LO) contribution to the next to LO  (NLO) contribution
of the three-gluon vertex, and we will consider convergence of the entire
procedure by evaluating the string tension at  different DSE scale fixing points.

Within QCD, the DSE  for propagators  and vertex functions have
been studied in great depth, for review see~\cite{Huber:2007kc,Alkofer:2008bs}
and references therein. Relation of DSE to lattice results is discussed in\mycite{Fischer:2008uz}. An alternative related method of functional renormalization group has been discussed in\mycite{Pawlowski:2003hq}.  The relevant results on three-point functions
are seen in~\cite{Alkofer:2008dt}, and on quark propagator
in~\cite{Alkofer:2008bs}, the  question of confinement inherent alone
in DSE are discussed in~\cite{Alkofer:2008tt,Alkofer:2007qf,Alkofer:2007zb},
the uniqueness of the infrared (IR) scaling of Green functions established
and gluon propagator IR non-singularity strictly supported
in~\cite{Fischer:2006vf,Fischer:2009tn}, the IR universality established in~\cite{Alkofer:2004it}.

Below in section~\ref{essz} we describe the   ESSZ equations in a pure Yang--Mills theory
with an arbitrary propagator (form-factor). In section~\ref{dys} we present DSE and  our
 solution, our results are in agreement with  the standard state-of-the art calculations
of ghost and gluon propagators in Landau gauge. In section~\ref{res} we evaluate the ESSZ
truncated Wilson loop employing the  DSE propagators from section~\ref{dys} and check the significance of the NLO vertex correction.
In section~\ref{disc} we discuss the reasons why confining potential is not
observed either in pure-glue two-point sector of DSE, or ESSZ solely,
yet it is seen in the combination thereof.

\section{ESSZ Equation \label{essz}}

The Wilson loop
\beq\displaystyle \label{wl}
W(C)=\langle \tr\,\, \mathrm{Pexp}\left\{ \oint_C A_\mu(x)dx^\mu\right\}\rangle
\eeq
offers information about the behaviour of quarks in the theory, and the
quark-antiquark potential is
\beq
V(L)=-\lim_{T\to \infty}\frac{1}{T} \ln W(C_{T,L}),
\eeq
where $C_{T,L}$ is a rectangular Wilson loop  in the $(x^0,x^1)$ plane, with $T$ being loop temporal length, and $L$ loop spatial length, $T\gg L$.

The Wilson loop \req{wl}  can be represented in terms of a perturbative expansion, which can be found e.g. in the review\mycite{Makeenko:1996bk}. A set of Feynman rules for Wilson loops can be found in\mycite{Brandt:1981kf}, which will be of use to us below. Perturbative treatment of Wilson loops is not useful in the non-Abelian case, and especially in the present context,
 as it yields obviously wrong results for Yang--Mills theory, for which
it predicts a Coulomb-type potential\mycite{Makeenko:1996bk}.
A large-$N_c$ partial summation of ladder diagrams has been proposed
 in\mycite{Erickson:1999qv,Erickson:2000af} and performed for a
circular and a rectangular loop in $\mathcal{N}=4$ supersymmetric model (SUSY).
This method is adapted here  to the case of a non-SUSY theory, for the case that
the partial summation of perturbation theory (PT) series for propagators
has already been performed in terms of solving DSE.

\begin{figure}[tbh]
\begin{center}
\includegraphics[height = 3.5cm, width=7cm]{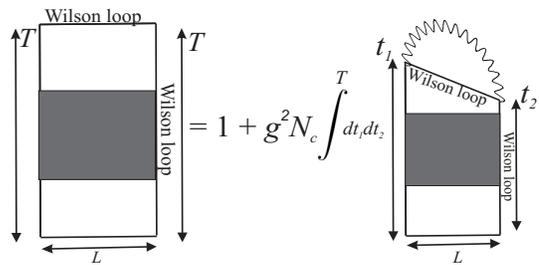}
\caption{Summation of ladder/rainbows for a Wilson loop. \label{resum}}
\end{center}
\end{figure}

Consider a trapezoidal loop $W(C)=\Gamma(T_1,T_2;L)$ with long parallel temporal sides of lengths $T_1,T_2$, separated by a spatial distance $L$. Then the requirement
that adding a propagator to the  summed expression does
not change it leads to the following integral equation for the sum of all ladder diagrams:
 \begin{eqnarray} \label{zarembo}
\Gamma(T_1,T_2,L)&=&1+ \frac{g^2N_c}{4\pi^2}\int^{T_1}dt_1\int^{T_2}dt_2\times\\[0.2cm] &&\times\Gamma(t_1,t_2,L)D_{\mu\nu}((x_1-x_2)^2)
\dot{x}_1^\mu\dot{x}_2^\nu,\notag
\end{eqnarray}
dots denote derivatives in $t_{1,2}$ respectively, where
 $x_1^\mu=x_1^\mu(t_1),\ \,x_2^\nu=x_2^\nu(t_1)$
are paths  running over the Wilson loop as functions of $t_1, t_2$. For a rectangular loop $x_1=(-L/2,t_1,0,0)$,$x_2=(L/2,t_2,0,0)$.  Configuration space propagator is related to the momentum-space form-factor $F(p^2)$, introduced in the next Section \ref{dys} by:
\beq
D_{\mu\nu}(x^2)=\frac{1}{(2\pi)^4}\int \frac{d^4p e^{-ipx}}{p^2}F(p^2)\left(g_{\mu\nu}-\frac{p_\mu p\nu}{p^2}\right).
\eeq
For simplicity we write $D_{\mu\nu}(x^2)
\dot{x}_1^\mu\dot{x}_2^\nu\equiv D(x^2)$.
Boundary conditions imposed upon $\Gamma$ are
\beq
\Gamma(T,0;L)=\Gamma(0,T;L)=1.
\eeq
The potential is related to $\Gamma(T_1,T_2;L)$
in the following way
\beq
V(L)=-\lim_{T\to\infty}\frac{1}{T}\ln \Gamma\left(T,T;L\right).
\eeq
Equation (\ref{zarembo}) is depicted symbolically in\myfigref{resum}.
 Obviously, if we write down the first
term for $\Gamma(T_1,T_2;L)$ in the $g^2$ expansion
of the solution, we shall reproduce the perturbative
result for the Wilson loop.

The central filled square in\myfigref{resum} symbolizes an irreducible kernel,
containing (potentially) all the possible loop corrections.
A convenient way of solving \req{zarembo} is   to consider the equivalent
differential equation:
\beq
\frac{\partial^2 \Gamma(t_1,t_2;L)}{\partial t_1\partial t_2}=\frac{g^2 N_c}{4\pi^2}D\left((t_1-t_2)^2+
L^2\right)\Gamma(t_1,t_2;L),
\eeq
We now introduce the variables $x=(t_1-t_2)/L,\,\, y=(t_1+t_2)/L$.
With this Ansatz the separation of variables becomes possible, and using the form:
\beq
\Gamma=\sum_n\psi_n(x)e^{\frac{\Omega_n y}{2 L}}
\eeq
we will be solving the 1d-equation
\beq\label{eric}
-\frac{d^2}{dx^2}\psi_n(x)+U(x;L)\psi_n(x)
=-\frac{\Omega_n^2}{4} \psi_n(x),
\eeq
with the effective potential
\beq\label{ericU}
U(x;L)=- \frac{g^2N_c}{4\pi^2}{L^2} D\left(L^2(1+x^2)\right)
\eeq
We are solely interested in the unique ground state solution of \req{eric}, since
the Wilson quark-quark potential  is
\beq
V(L)=-\lim_{T\to\infty}\frac{1}{T} \log \sum_n \psi_n(x) e^{\frac{\Omega_n T}{L}}=-\frac{\Omega_0}{L}.
\eeq
A degeneracy in solutions of \req{eric} may  arise and thus complicate the situation,
however, we have never observed it in our numeric calculations shown below
in section \ref{res}. It is now evident, that in order
to complete the Wilson potential evaluation
 we need the propagator $D$ and the coupling $\alpha=g^2/4\pi$ derived from DSE in
order to be able to  evaluate $V= {\Omega_0(L)}/{L}$.

\section{Dyson--Schwinger equations\label{dys}}
We now obtain the nonperturbative input to ESSZ equations, i.e.
the  Dyson--Schwinger improved gluon propagator and coupling $\alpha$.
The difference between DSE  and the simple
renormalization group (RG) improved
quantity is in the IR and medium momentum ranges,
their ultra violet (UV) behaviour being identical (up to 1 loop at least).
Our DSE procedure uses the technique  described in\mycite{Fischer:2002eq,Fischer:2002hna},
the reader familiar with this may skip the current section where we demonstrate
 that the results of\mycite{Fischer:2002eq,Fischer:2002hna} are independently reproduced by us.

\begin{figure}[h!] \begin{center}
\includegraphics[height = 3.cm, width=8cm]{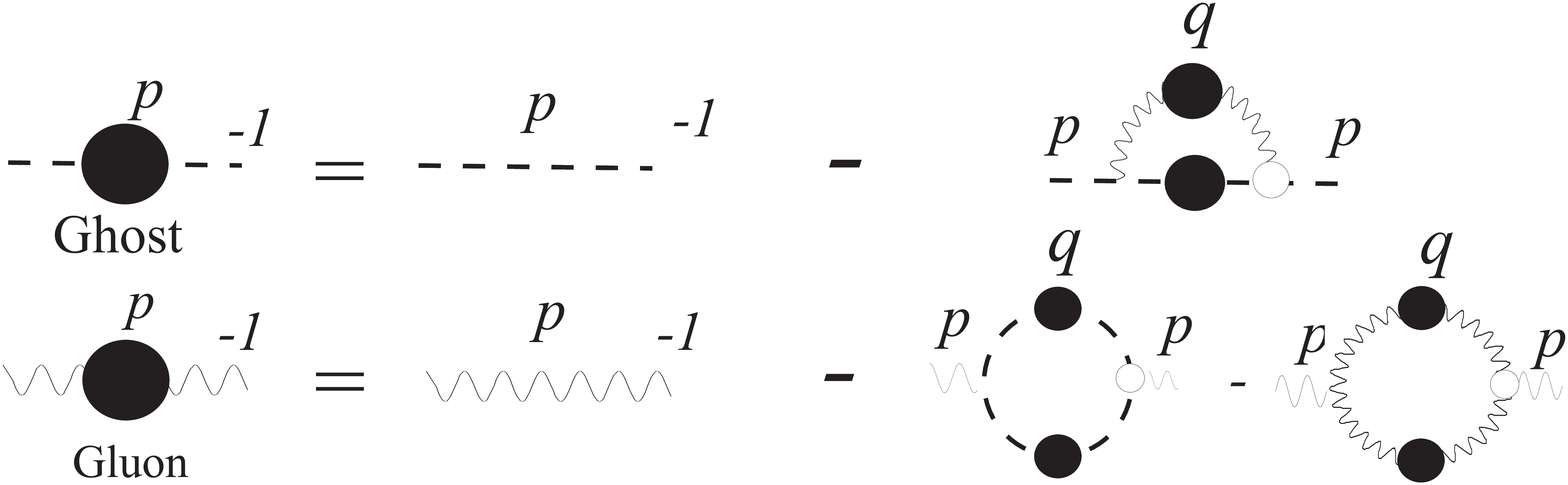}
\caption{Diagrammatic representation of DSE.
\label{dyson-diagr} }
\end{center}
\end{figure}

We employ in our work the Newton-method based numerical technique described
 in\mycite{Bloch:2003yu}.  We solve a system for ghost and gluon propagators,
corresponding to the representation seen in figure\myfigref{dyson-diagr}. Here
bulbs denote dressing of the propagators, and transparent bulbs -- dressing of vertices.

These equations can be written in the form:
\beq\displaystyle\label{dse}
\left\{\begin{array}{l}\displaystyle
\frac{1}{G(p^2)}-\frac{1}{G(\bar{\mu}_c^2)}=-\left(\Sigma(p^2)
-\Sigma(\bar{\mu}_c^2)\right),\\ \\ \displaystyle
\frac{1}{F(p^2)}-\frac{1}{F(\bar{\mu}_g^2)}=-\left(\Pi(p^2)
-\Pi(\bar{\mu}_g^2)\right),
\end{array}\right.
\eeq
where vacuum polarization is
\beq
\Pi(p^2)=\Pi^{2c}(p^2)+\Pi^{2g}(p^2),
\eeq

\beq
\begin{array}{l}\displaystyle
\Pi^{2c}(p^2)=N_c g^2\int  \frac{d^d q}{(2\pi)^d} M_0(p^2,q^2,r^2)G(q^2)G(r^2),\\  \\ \displaystyle
\Pi^{2g}(p^2)=N_c g^2\int  \frac{d^d q}{(2\pi)^4} Q_0(p^2,q^2,r^2)F(q^2)F(r^2),
\end{array}
\eeq
and self-energy is
\beq
\Sigma(p^2)=N_c g^2\int K_0(p^2,q^2,r^2)G(q^2) F(r^2) \frac{d^d q}{(2\pi)^d}.
\eeq
Here $\bar{\mu}_{g,c}$ are subtraction points, $\bar{\mu}_c=0$, $\bar{\mu}_g=\bar{\mu}$, $\bar{\mu}$ is the limit of the interval $p^2\in (0,\bar{\mu}^2)$ in the momentum space where we solve DSE, coupling $g^2$ is meant to be $g^2(\bar{\mu}^2)$. $F$ is gluon propagator form-factor in Landau gauge, defined via relation
\beq
\displaystyle
D_{\mu\nu}^{F\,ab}(p) =
\delta^{ab}\left(g_{\mu\nu}-\frac{p_\mu p_\nu}{p^2}\right)\frac{ F(p^2)}{p^2+i\epsilon},
\eeq
and the ghost propagator non-trivial behaviour is described by the form-factor $G$
\beq
\displaystyle
D^{G\,ab}(p)=\frac{\delta^{ab}}
{p^2+i\epsilon}{ G(p^2)}.
\eeq
Variable $z$ is the logarithmic variable
\beq
z=\ln\frac{p^2}{\mu^2},
\eeq
and scale $\mu$ is yet to be defined upon solving Dyson--Schwinger equations from comparing the obtained coupling $\alpha_{DSE}(z)$ to the known values of $\alpha_{PDG}(p^2)$ at point $M$:
\beq\label{normscale}
\alpha_{DSE}(\ln(M^2/\mu^2))=\alpha_{PDG}(M^2).
\eeq
The coupling constant $g^2/4\pi\equiv \alpha$
is expressed in terms of $G,F$ solely\mycite{von Smekal:1997is,von Smekal:1997vx}, as vertex
is finite in Landau gauge (at one-loop level)
\beq\label{alphDSE}
\alpha_{\mathrm{DSE}}(\ln(p^2))=\alpha_{\mathrm{DSE}}(\sigma)F(p^2)G^2(p^2).
\eeq
In our case, we shall use varying scale fixing  point $M$
so that we can prove that our results are independent
of scale fixing point choice within the error margin of our procedure.

The kernels $M_0,K_0,Q_0$ are known in literature, but for self-containedness of the paper we show them here:
\beq\displaystyle
\begin{array}{rcl}
\displaystyle
K_0(x,y,\theta)&=&\frac{y^2 \sin ^4(\theta )}
    {\left(-2 \cos (\theta ) \sqrt{x y}+x+y\right)^2},\\[0.4cm]
\displaystyle
M_0(x,y,\theta)&=&-\frac{y^2 \sin ^4(\theta )}
     {3 x \left(-2 \cos (\theta ) \sqrt{x y}+x+y\right)},
\end{array}
\eeq

\beq\displaystyle
\begin{array}{l}\displaystyle
Q_0(x,y,\theta)=-\frac{1}{12 x \left(-2 \cos (\theta ) \sqrt{x y}+x+y\right)^2}\times \\[0.4cm]
  \left\{y \sin ^2(\theta ) \left[2 \cos (2 \theta )
\left(6 x^2+31 x y+6 y^2\right)-\right.\right. \\[0.4cm]
-12 x \cos (3 \theta )
   \sqrt{x y}+x y \cos (4 \theta )
-48 \cos (\theta ) \sqrt{x y} (x+y)-\\ [0.4cm]
  \left.\left.
-12 y \cos (3 \theta ) \sqrt{x y}+3
   x^2+27 x y+3 y^2\right]\right\}.
\end{array}
\eeq
For convenience, variables $x=p^2$, $y=q^2$ are introduced; variable $\theta$ is defined via $(p-q)^2=x+y-2\sqrt{xy}\cos\theta$.

\begin{figure}[tbh]
\begin{center}
\includegraphics[height = 4.5cm, width=8cm]{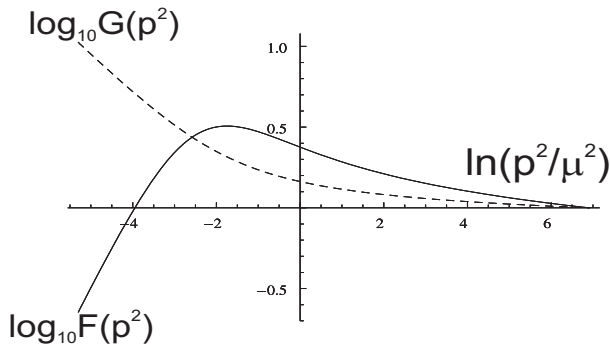}
\caption{Ghost (dashed line) and gluon (solid line) propagator formfactors obtained in
DSE in Landau gauge.
\label{prop}}
\end{center}
\end{figure}

To solve Dyson--Schwinger equations we use the Ansatz\mycite{Fischer:2002eq,Fischer:2002hna}:
\beq
\begin{array}{l}\displaystyle
F(z)=\left\{
\begin{array}{l}\displaystyle
\exp\left(\sum^{\bar{n}}_i a_i T_i(z)\right),\, z\in(\ln\epsilon,\ln\bar{\mu}^2),\\ \\ \displaystyle
F(\bar{\mu})\left(1+
\omega\log\frac{p^2}{\bar{\mu}^2}\right)^\gamma,
z>\ln\bar{\mu}^2,
  \\ \\ \displaystyle
A z^{2\kappa},z<\ln \epsilon,
\end{array}\right.\\  \\ \displaystyle
G(z)=\left\{
\begin{array}{l}\displaystyle
\exp\left(\sum^{\bar{n}}_i b_i T_i(z)\right),\,
   z\in(\ln\epsilon,\ln\bar{\mu}^2),\\  \\ \displaystyle
G(\sigma)\left(1+
\omega\ln\frac{p^2}{\bar{\mu}^2}
\right)^\delta,z>\ln\bar{\mu}^2,\\  \\ \displaystyle
B z^{-\kappa}, z<\epsilon.
\end{array}\right.\\
\end{array}
\eeq
Here $T_i$ are Tschebyschev polynomials, $a_i,b_i$
are unknown coefficients yet to be determined from
the numerical solution,  $\bar{n}$ is the number
of polynomials used (mostly $\bar{n}=30$ has been
used here, allowing precision of $10^{-10}$ for
coefficients), $\delta=-9/44$,
$\gamma=-1-2\delta$, $\omega=11 N_c \alpha(\sigma)/(12\pi)$.
 The IR scaling $\kappa$ is chosen to be the standard\mycite{Lerche:2002ep,Zwanziger:2001kw}
\beq
\kappa=0.59\label{kappa}
\eeq
for the case of Brown--Pennington
truncation with $\zeta=1$\mycite{Fischer:2002hna} (for discussion of meaning of $\zeta$ see \mycite{Brown:1988bm}), which is our case
($\zeta$ already set to its number value everywhere).
 Following\mycite{Alkofer:2008bs}, we employ renormalization
 constant $\mathcal{Z}_1$ redefinition,
so that no momentum dependence could possibly enter it, that is
\beq
\mathcal{Z}_1=\frac{G(y)^{(1-a/\delta -2a)}}{F(y)^{(1+a)}}
 \frac{G(y)^{(1-b/\delta -2b)}}{F(y)^{(1+b)}}.
\eeq
Again, following\mycite{Alkofer:2008bs} we choose
\beq
a=b=3\delta,
\eeq
which minimizes its momentum dependence. Renormalization
constant $\mathcal{Z}_1$ refers to the piece with ghost
loop in vacuum polarization. The equations are solved by using
Newton's method, very clearly described for this particular
application by Bloch\mycite{Bloch:2003yu}.
 The results of the solution are propagator formfactors
$F,G$, shown in\myfigref{prop},
the IR behaviour of the propagators
corresponds to the standard ghost enhancement and gluon suppression.

\begin{figure}[tbh]
\begin{center}
\includegraphics[height = 4.5cm, width=8cm]{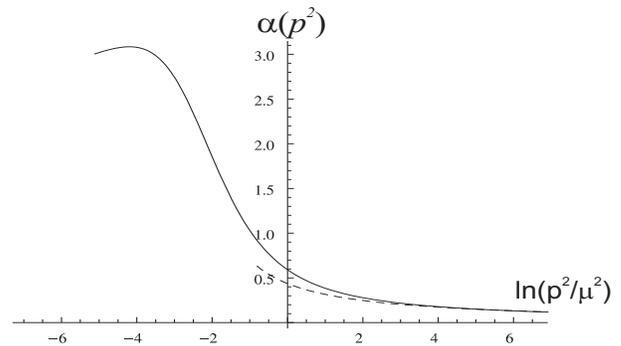}
\caption{Running coupling from Dyson--Schwinger equations: our solution of DSE (solid line), and standard PDG coupling (dashed).
\label{alpha}}
 \end{center}
\end{figure}

The coupling $\alpha$ obtained from DSE \req{alphDSE}  is shown in\myfigref{alpha}.
We compare it to the standard coupling from Particle Data Group~\cite{Amsler:2008zzb},
 and note that the both coincide very well in the UV. We also note here that
the IR fixed point seen in the Figure is
\beq
\alpha(0)\approx 2.9 \notag
\eeq
for $N_c=3$, which is consistent with the up-to-date Dyson--Schwinger results reported
by other groups\mycite{Alkofer:2008bs,Huber:2007kc}.

\section{Solving ESSZ Equation\label{res}}
\begin{figure}[tbh]
\begin{center}
\includegraphics[height = 4cm, width=8cm]{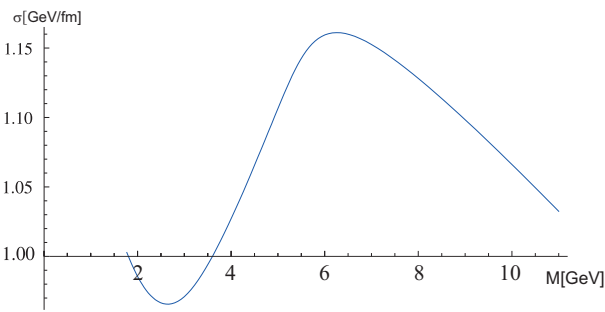}
\caption{Dependence of string tension $\sigma$ on scale fixing point $M$.
\label{sigma}} \end{center}
\end{figure}

We have to find the lowest eigenvalue of a Schr\"odinger equation \req{eric}
\beq
\left(-\frac{1}{2}\frac{d^2}{dx^2}
+ U(x;L)\right)
\psi(x)=\mathcal{E}\psi(x)
\eeq
where the auxillary potential $U(x)$ is related by an linear integral transform to gluon form-factor as
\beq\displaystyle
\begin{array}{l}\displaystyle
U(x)=-\frac{2\pi \alpha N_c}{(1+x^2)^2}\frac{1}{(2\pi)^2}\int \frac{du}{u}
\times\\[0.4cm]  \displaystyle
\left(uJ_1(u)-(1-3x^2)J_2(u)\right)
F\left(\ln\left(\frac{u^2}{L^2 \mu^2 (1+x^2)}\right)\right),
\end{array}
\eeq
where $\mu$ is defined at point $M$ as given in\myref{normscale}, $M$ varying from 1 to 10 GeV, $u$ is dummy scalar dimensionless integration variable. The coupling $\alpha$, in the sense of DSE approach, is taken here at the scale of $\frac{2\pi}{L}$, rather than bare. We solve the Schr\"odinger equation with shooting method and find its ground state. Special care is taken to make sure this state is not degenerate. As a result we get the QCD potential $V(L)=-\frac{2\sqrt{2|\mathcal{E}|}}{L}$. The potential is defined up to additive constant, so we shift it to provide convenient comparison to existent results. It is shown in\myfigref{experiment} below, and is compared with lattice results by Gubarev et al.\mycite{Boyko:2007jx}  and Necco\mycite{Necco:2003jf}. Linear IR behaviour of the potential can be clearly seen from Figure. We fit the potential by the standard expression
\beq
V(L)=-\frac{4}{3}\frac{\alpha_0}{L}+c_0+\sigma L.
\eeq
Dependence on string tension $\sigma$ on the scale fixing point choice is shown in\myfigref{sigma}. We see that the variance of $\sigma$ does not exceed that of different lattice results, shown in the table\myref{complatt}.
The error we quote arises from  an average of results obtained
at different scale fixing points. This yields $\alpha_0=0.24$ and $\sigma=1.07\pm0.1$.

\begin{table}
\caption{Comparison of string tension from different sources\label{complatt}}
\begin{center}
\begin{tabular}{|l|l|l|}\hline
Author&Year&$\sigma$,GeV/fm\\ \hline
Bali et al.\mycite{Bali:2000vr}&2000&1.27\\ \hline
Necco\mycite{Necco:2003jf}&2003&1.19\\ \hline
Gubarev et al.\mycite{Boyko:2007jx}&2007&0.978\\ \hline
Weise et al.\mycite{Laschka:2009um}&2009&1.07\\ \hline
Present work&2009&$1.07\pm 0.1$\\ \hline
\end{tabular}
\end{center}
\end{table}

\begin{figure}[tbh]
\begin{center}
\includegraphics[height = 4.cm, width=8cm]{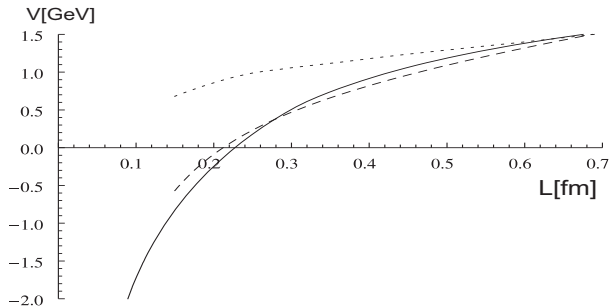}
\caption{Potential  as function of distance,
solid line -- our result,
dashed line -- result by Necco, 2003\mycite{Necco:2003jf},
dotted line -- result by Gubarev et al. 2007\mycite{Boyko:2007jx}.
 \label{experiment}} \end{center}
\end{figure}

\begin{figure}[thb]
\begin{center}
\includegraphics[height = 6cm, width=6cm]{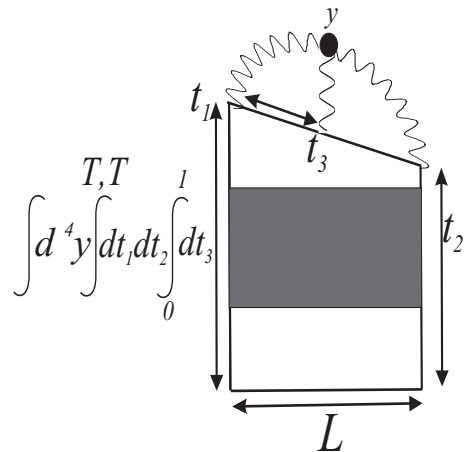}
\caption{Terms in ESSZ equations generating two-loop correction
to the auxillary potential $U^{(2)}(x)$.
\label{twoloop}}
\end{center}
\end{figure}

The key result, the linear confining potential
comes as a surprise. It invites the question,
how great are the corrections coming from three-point vertex?
One actually shouldn't have thought that QCD can be described with ESSZ
 partial summation structure. Considering the vertex,
the auxillary potential is then modified:
\beq
U(x)=U^{(1)}(x)+4\pi \alpha N_c U^{(2)}(x),
\eeq
where $U^{(2)}(x)$ comes, in the leading $1/N_c$ order,
from the Wilson loop diagram shown in\myfigref{twoloop}.

Calculating the diagram in Landau gauge with rules  as defined in\mycite{Brandt:1981kf}
we obtain:
\beq
\begin{array}{l}\displaystyle
U^{(2)}(t_1,t_2)=\int d^4y \int^{1}_0 dt_3 \frac{1}{(y-x_1)^2 (y-x_2)^2 (y-x_3)^2}\times
\\[0.4cm] \displaystyle
\left[(u_1 u_2) (u_3 y) \left(\frac{1}{(y-x_1)^2}-\frac{1}{(y-x_2)^2}\right)
+\mbox{cyclic permut.}\right]
\end{array}
\eeq
with
\beq
\left\{\begin{array}{l}
x_1=(-L/2,t_1,0,0)\\
x_2=(L/2,t_2,0,0)\\
x_3=(-L/2+Lt_3,t_1+t_3(t_2-t_1),0,0)\\
u_1=(0,1,0,0)\\
u_2=(0,1,0,0)\\
u_3=(L,t_2-t_1,0,0)\\
\end{array}\right.
\eeq
Numerical evaluation of this integral shows that within the
whole range of values of $t_1,t_2$ with which we work,
$U^{(2)}(t_1,t_2)=U^{(2)}(t_1-t_2)\equiv U^{(2)}(x)$.
This makes separation of variables still possible and
provides an extra test for reasonability of our model.
Numerical values of $U^{(2)}(x)$ are such that
$U^{(2)}(x)/U^{(1)}(x)\stackrel{\large <}{\approx} 10^{-3}$,
which makes its contribution to auxillary potential
ground state negligible. This allows us to justify
validity of ESSZ equation application in the non-SUSY
 case: vertex correction is present but numerically suppressed.

\section{Discussion and Conclusion \label{disc}}

ESSZ approach to SUSY Wilson loops has worked very well in\mycite{Erickson:1999qv,Erickson:2000af}.
 The reason for that is absence of NLO corrections in
the maximally supersymmetric theory. At small coupling
their result has restored the perturbatively known
IR singularity structure. Moreover, the calculation
originally performed in the small coupling limit,
could  be continued into large coupling limit.
At large coupling the solution to ESSZ equation
reproduces almost exactly the gravity dual
result~\mycite{Maldacena:1998im,Rey:1998ik,Brandhuber:1998er}
(up to an overall numerical factor very close to unity). Actually this result, though in a different theory, has been a guide for our QCD treatment: as we are dealing with the IR strongly coupled theory, we are certainly out of order of applicability of any perturbative treatment, and even summation of diagrams would be suspicious.

The reason why the ESSZ equation has never been applied
to non-SUSY contents is obvious. It is clear
from\mycite{Erickson:1999qv, Erickson:2000af} that
when a perturbative propagator input is being used
only a non-confining Wilson loop, with a Coulomb-type
potential may be obtained. This follows from the
 fact that dependence on Wilson loop spatial
size $L$ may be scaled out of the ESSZ equation,
so that any potentials one gets from it are Coulombic,
varying from each other by coupling rather than distance
dependence. Thus such a result would have been
{\it a priori} useless in understanding anything
about strong coupling IR regime of gauge theory,
where confinement governs the dynamics. This maybe
the reason  why summation \`a la ESSZ has not
before  been employed in pure Yang--Mills theory.

A direct
perturbative calculation of Wilson loop with
a Dyson--Schwinger propagator yields no confinement
 whatsoever. Only quark-gluon vertex
functions\mycite{Alkofer:2008tt,
Alkofer:2007qf,Alkofer:2007zb}
coming from DSE can render something looking like confinement,
which is then related to the singular behaviour of the quark-gluon
vertex in the IR.

In our opinion, this could not be one of the possible ways to approach within the DSE
the confinement  problem, since it requires quark coupling to be singular, gluon one regular, which constitutes a severe violation of Ward identities. Rather than to involve three-particle functions,
we apply ESSZ summation with DSE solutions which are   possessing
intrinsic scale, distance $L$ is no more possible to scale
 out of ESSZ equations. Thus the resulting Wilson
potential is no more necessarily being Coulombic.

A description of a single Wilson loop, from which one can obtain the QCD potential and provide a criterion of confinement, has not been done so far in terms of the two-point sector of DSE hierarchy. Thus our work closes an essential gap in the literature. The main reason for this gap was the
theorem by West\mycite{West:1982bt}, stating that confinement is
provided by a very IR-singular propagator
$D(q^2)\sim 1/q^4$. We know however
that gluon propagator is regular in the IR
in the DSE approach.

Our work is based on combined analysis of Green functions and Wilson loops, allowing thus a study of the spatial QCD potential. This distinguishes our approach from several earlier papers where gluon non-propagation was considered instead of confinement and related to the analytic properties of Green functions, in particular, to the IR scaling $\kappa$, Eq.\myref{kappa}. These other works use the word ``confinement'' as in the original paper\mycite{Wilson:1974sk} when they mean to say of ``non-propagation''. Known are the so-called Kugo--Ojima criterion for colour non-propagation $\kappa>0$\mycite{Kugo:1979gm}, Zwanziger criterion of ghost non-propagation $\kappa>0$ and gluon non-propagation
$\kappa>1/2$\mycite{Zwanziger:1991gz}. A claim has been made
\mycite{Braun:2007bx} for $\kappa>1/4$ to be quark confinement criterion by analysis of the Polyakov loop and effective QCD action in an external field. All these results are about gluon non-propagation rather than the properties of a colour charge-colour charge confining interaction.

Returning to the discussion of our results we note that the reliability thereof may be questioned in what concerns
the DSE input. The first issue is the truncation of the DSE system
we solve to only   two-point functions. The truncation is justified by ghost-gluon vertex not acquiring acquire one-loop corrections in Landau gauge. It has been proved that the three-point gluon and quark-gluon functions don't
change ghost dominance property\mycite{Alkofer:2008dt}, even though they are important for bound states\mycite{Cucchieri:2008qm}.
In this sense, vertex functions are unimportant for our
particular context. 

Another question is whether Green
functions obtained from DSE are physically relevant within the Wilson loop context we are discussing.
We note that it is mostly medium-energy range that provides
the important contribution into the auxillary
potential $U(x;L)$, rather than the perhaps more model dependent
IR piece. The Wilson loop thus depends on medium energy range values of the propagators where the DSE behaviour is the same as in lattice. There are unresolved questions regarding comparison of IR scaling\mycite{Sternbeck:2007ug} within lattice and DSE. These issues have yet to be understood and resolved, although they do not affect our results materially.

The observables $\sigma,\alpha$ we compute are in principle gauge invariant. Our results are obtained in Landau gauge, which, as noted, is a convenient choice. It should be possible to check gauge-invariance explicitly at one loop level, we however do not do that here, since this transcends the scope of the present paper. We think that the possibility for the observable we consider to be gauge invariant at one-loop level comes from the fact that several gauge-dependent objects are combined.

We speculate here {\it \`a propos} that a nonperturbative  summation
{\it a la} ESSZ could
improve significantly the properties of a correlator
of gluon strengths with Wilson lines
\beq\label{Simonov}
\mathcal{F}(x)=\langle\tr F_{\mu\nu}(x)U(C)F^{\mu\nu}(0)U^+(C)\rangle
\eeq
$U(C)$ being a   phase factor
\beq
U(C)=\mathrm{Pexp}\left\{ig\int_C A_\mu dx^\mu\right\},
\eeq
which differs from Wilson loop since the path is connecting the
arguments  in \req{Simonov} i.e. points $x$ and $0$. \req{Simonov}
 hade recently been of great interest\mycite{Di Giacomo:2005cv}
as it represents an  important vacuum property.
As far as we know, a Bethe--Salpeter equation for this kind of
correlator has not been developed  yet. We  attempted
to  evaluate it perturbatively\mycite{Zayakin:2008uq}.
The present effort arose from this earlier one but
should have actually anteceded it, for then a
framework for ESSZ summation may have been closer or even in
hand.

A  hypothesis should be considered  that
using a relevant component  of the non-perturbative input from
Dyson--Schwinger equations one may be able to obtain
a self-consistent picture of the QCD vacuum with
 all higher correlation functions, colour
confinement and condensates, which is
supported at the simplest LO level
by the presented calculation.

To conclude,
combining Dyson--Schwinger summation for gluon and ghost propagators
with the Ericson--Semenoff--Szabo--Zarembo summation (truncation)
 for Wilson loop, we have  obtained the string tension
 and have further demonstrated that  its value  is nearly not dependent
on the selection of the DSE scale fixing point, thus establishing
the internal consistency of this novel description of confinement. The string tension determined by our method for the pure SU(3)
gauge theory is $\sigma=1.07\pm 0.1$. The UV Couloumb behaviour is governed by $\alpha_0\approx 0.24$.

One can actually be quite amazed that our method has worked
so well in QCD, without supersymmetry, thus with vertices non-compensated.
One can speculate  that the two truncated summations  are complementary,
 ESSZ taking care of ladders and DSE taking care of rainbows
in the vertices. Among interesting  further steps
in the development of this framework we recognize the
formulation and evaluation of a similar ESSZ equation
or a correlator of two gluons, having
in mind its application to gluon non-local condensate \req{Simonov}.
Another, perhaps more challenging further development could be
to solve ESSZ and DSE  jointly, without the separation
into partial systems.


\begin{acknowledgments}
We thank Prof. Dr. D. Habs  for   hospitality
at the Physics Department at LMU Garching.
One of us (A.Z.) thanks K.Zarembo for useful correspondence.
This work was supported by the DFG Cluster of Excellence MAP
(Munich Centre of Advanced Photonics), by RFBR Grant 07-01-00526,
and by a grant from the U.S. Department of Energy  DE-FG02-04ER4131.
\end{acknowledgments}


\begin{thebibliography}{99}


\bibitem{Wilson:1974sk}
  K.~G.~Wilson,
  Phys.\ Rev.\  D {\bf 10}, 2445 (1974).

\bibitem{Erickson:1999qv}
  J.~K.~Erickson, G.~W.~Semenoff, R.~J.~Szabo and K.~Zarembo,
  Phys.\ Rev.\  D {\bf 61}, 105006 (2000)
  [arXiv:hep-th/9911088].


\bibitem{Erickson:2000af}
  J.~K.~Erickson, G.~W.~Semenoff and K.~Zarembo,
  Nucl.\ Phys.\  B {\bf 582}, 155 (2000)
  [arXiv:hep-th/0003055].








\bibitem{Huber:2007kc}
  M.~Q.~Huber, R.~Alkofer, C.~S.~Fischer and K.~Schwenzer,
  Phys.\ Lett.\  B {\bf 659}, 434 (2008)
  [arXiv:0705.3809 [hep-ph]].

\bibitem{Alkofer:2008bs}
  R.~Alkofer, C.~S.~Fischer, M.~Q.~Huber, F.~J.~Llanes-Estrada and K.~Schwenzer,
  arXiv:0812.2896 [hep-ph].

\bibitem{Fischer:2008uz}
  C.~S.~Fischer, A.~Maas and J.~M.~Pawlowski,
  arXiv:0810.1987 [hep-ph].
  
\bibitem{Pawlowski:2003hq}
  J.~M.~Pawlowski, D.~F.~Litim, S.~Nedelko and L.~von Smekal,
  Phys.\ Rev.\ Lett.\  {\bf 93}, 152002 (2004)
  [arXiv:hep-th/0312324].


\bibitem{Alkofer:2008dt}
  R.~Alkofer, M.~Q.~Huber and K.~Schwenzer,
  arXiv:0812.4045 [hep-ph].



\bibitem{Alkofer:2008tt}
  R.~Alkofer, C.~S.~Fischer, F.~J.~Llanes-Estrada and K.~Schwenzer,
  Annals Phys.\  {\bf 324}, 106 (2009)
  [arXiv:0804.3042 [hep-ph]].

\bibitem{Alkofer:2007qf}
  R.~Alkofer, C.~S.~Fischer, F.~J.~Llanes-Estrada and K.~Schwenzer,
  PoS {\bf LAT2007}, 286 (2007)
  [arXiv:0710.1154 [hep-ph]].

\bibitem{Alkofer:2007zb}
  R.~Alkofer, C.~S.~Fischer, F.~J.~Llanes-Estrada and K.~Schwenzer,
  Int.\ J.\ Mod.\ Phys.\  E {\bf 16}, 2720 (2007)
  [arXiv:0705.4402 [hep-ph]].

\bibitem{Fischer:2006vf}
   C.~S.~Fischer and J.~M.~Pawlowski,
   Phys.\ Rev.\  D {\bf 75}, 025012 (2007)
   [arXiv:hep-th/0609009].

\bibitem{Fischer:2009tn}
  C.~S.~Fischer and J.~M.~Pawlowski,
  arXiv:0903.2193 [hep-th].


\bibitem{Alkofer:2004it}
   R.~Alkofer, C.~S.~Fischer and F.~J.~Llanes-Estrada,
   Phys.\ Lett.\  B {\bf 611}, 279 (2005)
   [Erratum-ibid.\  {\bf 670}, 460 (2009)]
   [arXiv:hep-th/0412330].



\bibitem{Makeenko:1996bk}
  Y.~M.~Makeenko,
  Surveys High Energ.\ Phys.\  {\bf 10}, 1 (1997).

\bibitem{Brandt:1981kf}
  R.~A.~Brandt, F.~Neri and M.~a.~Sato,
  Phys.\ Rev.\  D {\bf 24}, 879 (1981).

\bibitem{von Smekal:1997is}
  L.~von Smekal, R.~Alkofer and A.~Hauck,
  Phys.\ Rev.\ Lett.\  {\bf 79}, 3591 (1997)
  [arXiv:hep-ph/9705242].

\bibitem{von Smekal:1997vx}
  L.~von Smekal, A.~Hauck and R.~Alkofer,
  Annals Phys.\  {\bf 267}, 1 (1998)
  [Erratum-ibid.\  {\bf 269}, 182 (1998)]
  [arXiv:hep-ph/9707327].



\bibitem{Fischer:2002eq}
  C.~S.~Fischer, R.~Alkofer and H.~Reinhardt,
  Phys.\ Rev.\  D {\bf 65}, 094008 (2002)
  [arXiv:hep-ph/0202195].

\bibitem{Fischer:2002hna}
   C.~S.~Fischer and R.~Alkofer,
   Phys.\ Lett.\  B {\bf 536} (2002) 177
   [arXiv:hep-ph/0202202].


\bibitem{Bloch:2003yu}
  J.~C.~R.~Bloch,
  Few Body Syst.\  {\bf 33}, 111 (2003)
  [arXiv:hep-ph/0303125].

\bibitem{Lerche:2002ep}
   C.~Lerche and L.~von Smekal,
   Phys.\ Rev.\  D {\bf 65}, 125006 (2002)
   [arXiv:hep-ph/0202194].

\bibitem{Zwanziger:2001kw}
   D.~Zwanziger,
   Phys.\ Rev.\  D {\bf 65}, 094039 (2002)
   [arXiv:hep-th/0109224].

\bibitem{Brown:1988bm}
  N.~Brown and M.~R.~Pennington,
  Phys.\ Rev.\  D {\bf 38}, 2266 (1988).


\bibitem{Amsler:2008zzb}
  C.~Amsler {\it et al.}  [Particle Data Group],
  Phys.\ Lett.\  B {\bf 667}, 1 (2008).






  \bibitem{Boyko:2007jx}
  P.~Y.~Boyko, F.~V.~Gubarev and S.~M.~Morozov,
  PoS {\bf LAT2007}, 307 (2007)
  [arXiv:0712.0656 [hep-lat]].

\bibitem{Necco:2003jf}
  S.~Necco,
  arXiv:hep-lat/0306005.



\bibitem{Bali:2000vr}
  G.~S.~Bali {\it et al.}  [TXL Collaboration and T(X)L Collaboration],
  Phys.\ Rev.\  D {\bf 62}, 054503 (2000)
  [arXiv:hep-lat/0003012].


\bibitem{Laschka:2009um}
  A.~Laschka, N.~Kaiser and W.~Weise,
  arXiv:0901.2260 [hep-ph].
  
  
\bibitem{Kugo:1979gm}
  T.~Kugo and I.~Ojima,
  Prog.\ Theor.\ Phys.\ Suppl.\  {\bf 66}, 1 (1979).
  
\bibitem{Zwanziger:1991gz}
  D.~Zwanziger,
  Nucl.\ Phys.\  B {\bf 364} (1991) 127.

\bibitem{Braun:2007bx}
  J.~Braun, H.~Gies and J.~M.~Pawlowski,
  arXiv:0708.2413 [hep-th].

\bibitem{West:1982bt}
  G.~B.~West,
  Phys.\ Lett.\  B {\bf 115}, 468 (1982).






\bibitem{Maldacena:1998im}
  J.~M.~Maldacena,
  Phys.\ Rev.\ Lett.\  {\bf 80}, 4859 (1998)
  [arXiv:hep-th/9803002].

\bibitem{Rey:1998ik}
  S.~J.~Rey and J.~T.~Yee,
  Eur.\ Phys.\ J.\  C {\bf 22}, 379 (2001)
  [arXiv:hep-th/9803001].


\bibitem{Brandhuber:1998er}
  A.~Brandhuber, N.~Itzhaki, J.~Sonnenschein and S.~Yankielowicz,
  JHEP {\bf 9806}, 001 (1998)
  [arXiv:hep-th/9803263].

\bibitem{Cucchieri:2008qm}
  A.~Cucchieri, A.~Maas and T.~Mendes,
  Phys.\ Rev.\  D {\bf 77}, 094510 (2008)
  [arXiv:0803.1798 [hep-lat]].




\bibitem{Sternbeck:2007ug}
  A.~Sternbeck, L.~von Smekal, D.~B.~Leinweber and A.~G.~Williams,
  PoS {\bf LAT2007}, 340 (2007)
  [arXiv:0710.1982 [hep-lat]].


\bibitem{Di Giacomo:2005cv}
  A.~Di Giacomo, E.~Meggiolaro, Yu.~A.~Simonov and A.~I.~Veselov,
  Phys.\ Atom.\ Nucl.\  {\bf 70} (2007) 908
  [arXiv:hep-ph/0512125].




\bibitem{Zayakin:2008uq}
  A.~V.~Zayakin and J.~Rafelski,
  arXiv:0812.3616 [hep-ph].
   \end{thebibliography}
\end{document}